\documentclass{article}
\usepackage{glas}
\usepackage{psfig}  \usepackage{epsfig}  \usepackage{graphics}
\begin{document}
\begin{titlepage}{GLAS-PPE/1999--11}{August 1999}
\title{MLLA Parton Spectra Compared to ARIADNE}
\author{N.\ H.\ Brook}
\begin{abstract}
The parton spectra as predicted by the ARIADNE Monte Carlo
generator, for both $e^+e^-$ annihilation and deep inelastic
scattering, are compared to
the QCD MLLA calculations.
\end{abstract}
\vfill
\conference{contribution to the \\
 1998-1999 HERA Monte Carlo workshop.}
\end{titlepage}
%
\parskip 2mm plus 2mm minus 1mm     
\renewcommand{\topfraction}{1.0}    
\renewcommand{\bottomfraction}{1.0} 
\renewcommand{\textfraction}{0.0}   
%
\def\Journal#1#2#3#4{{#1} {\bf #2}, #3 (#4)}
\def\NCA{\em Nuovo Cimento}
\def\NIM{\em Nucl. Instrum. Methods}
\def\NIMA{{\em Nucl. Instrum. Methods} A}
\def\NPB{{\em Nucl. Phys.} B}
\def\PLB{{\em Phys. Lett.}  B}
\def\PRL{\em Phys. Rev. Lett.}
\def\PRD{{\em Phys. Rev.} D}
\def\ZPC{{\em Z. Phys.} C}
%
\newcommand{\Pom}{I$\!$P}                
\def\lsim{\mathrel{\rlap{\lower4pt\hbox{\hskip1pt$\sim$}}
    \raise1pt\hbox{$<$}}}                
\def\gsim{\mathrel{\rlap{\lower4pt\hbox{\hskip1pt$\sim$}}
    \raise1pt\hbox{$>$}}}                
\newcommand{\ee}{\mbox{$e^+e^-$}}
%

\section{Introduction}
The perturbative QCD approach to describing 
the inclusive energy spectra, via the modified leading log
approximation (MLLA) in conjunction with
 local parton hadron duality (LPHD), has been very
successful in both $e^+e^-$ annihilation and deep inelastic
scattering experiments~\cite{MLLArev}. 
Using LPHD, the non-perturbative effects of such
distributions are reduced to a simple factor of normalisation that
relates the hadronic distributions to the partonic ones. Perturbative
features of these distributions are calculated by MLLA which accounts
for both the double and single logarithmic effects. The MLLA
approach has two free parameters: a running strong coupling, governed by
a QCD scale $\Lambda,$ and an energy cut-off, $Q_0,$ below which the
parton evolution is truncated.

The MLLA evolution equation allow the parton spectra for the
logarithmic scaled energy spectra, $\xi,$ to be calculated~\cite{dokevol}. 
The variable $\xi$ is
defined as $\ln(E_0/E) \equiv \ln(1/x_p),$ 
where $E_0$ is the original energy of the jet
and $E$ is the parton's energy. The cut-off, $Q_0,$ bounds the parton
energy, $ E \ge k_T \ge Q_0,$ where $k_T$ is the transverse energy of
the decay products in the jet evolution. In order
to reconstruct the $\xi$ distributions
one has to perform the inverse Mellin transformation:

\begin{equation}
\bar D(\xi,Y,\lambda) = \int_{\epsilon - \imath\infty}^{\epsilon + \imath\infty}
\frac{d\omega}{2\pi \imath} x_p^{-\omega} D(\omega,Y,\lambda)
\label{eqn:MLLA}
\end{equation}
where the integral runs parallel to the imaginary axis on the right of
all singularities in the complex $\omega-$plane, $Y = \ln(E_0/Q_0)$
and $\lambda=\ln(Q_0/\Lambda).$

The Mellin-transformed distributions, $D(\omega,Y,\lambda),$ can be
expressed~\cite{dokevol}  
in terms of confluent hypergeometric functions, $\Phi$:

\begin{equation}
\begin{array}{lc}
D(\omega,Y,\lambda) &=  \frac{t_1A}{B(B+1)}\Phi(-A+B+1,B+2;-t_1)\Phi(A-B,1-B;t_2) \\
 &+\left(\frac{t_2}{t_1}\right)^B\Phi(-A,-B;-t_1)\Phi(A,B+1;t_2),
\end{array}
\end{equation}
where 
\begin{equation}
\begin{array}{lr}
t_1 = \omega(Y+\lambda) \;\;,&  \;\; t_2= \omega\lambda.
\end{array}
\end{equation}
In addition $A$ and $B$ are defined as:

\begin{equation}
\begin{array}{lr}
A=4N_c/b\omega,\;\; & \;\;B=a/b,
\end{array}
\end{equation}
where $N_c$ is the number of colours, $a=11N_c/3+2n_f/3N_c^2$ , $n_f$
is the number of flavours and $b = 11N_c/3 + 2n_f/3.$

Equation~\ref{eqn:MLLA} is then calculated using a numerical integration
in the complex $\omega-$plane.

The current region in the $ep$ Breit frame
is analogous to a single hemisphere of $e^+e^-$ annihilation.
In $e^+e^- \rightarrow q \bar q$ annihilation the two quarks are
produced
with equal and opposite momenta, $\pm \sqrt{s}/2,$
where $\sqrt s$ is the positron-electron centre of mass energy.
The fragmentation of these quarks can be compared to
that of the quark struck from the
proton; this quark has an outgoing momentum $-Q/2$ in the Breit frame,
where $Q^2$ is the negative square of the
four-momentum of the virtual exchanged boson in DIS.
In the direction of this struck quark
the scaled momentum spectra of the particles
are expected by MLLA~\cite{eedis,anis,char} to have a
dependence on $Q$ similar to that observed in
\ee~annihilation at energy $\sqrt{s}=Q,$ with no Bjorken-$x$ dependence.

The ARIADNE Monte Carlo generator~\cite{ARIADNE}
is based on the colour dipole model, CDM~\cite{CDM}.
In the CDM, all gluon emissions constituting the QCD cascade start as
radiation from the colour dipole formed between the quark and the
anti-quark in the case of
$e^+e^-$ annihilation or the struck quark and the proton
remnant in the case of DIS. 
All subsequent radiation arises from independent colour
dipoles formed either from $q\bar{q}$ pairs or softer gluons radiated by
the
previously produced gluons. 
In the DIS scenario, the proton
remnant is treated as an extended object which
results in a suppression of radiation~\cite{DIP}, generally in
the proton direction. 
In addition the struck quark is treated as extended, as the
photon only
probes it to a distance inversely proportional to the transferred
momentum.  
Treating the remnant and the struck quark as extended objects, rather
than point like, results in a reduction in the available phase space for
gluon radiation in DIS.

The QCD cascade in ARIADNE is governed by a number of parameters in the
Monte Carlo models. Two of the most important are the QCD scale,
$\Lambda,$ (PARA(1)) and the parameter that determines the $k_T$ cut off for the
shower (PARA(3)). An additional parameter, PARA(28), also allows the
user to bound the lower energy of the emitted parton as well.
For this study ARIADNE version 4.10 has been used.

\section{Comparisons with MLLA}
\begin{figure}[pt!]
\begin{center}\mbox{\epsfig{file=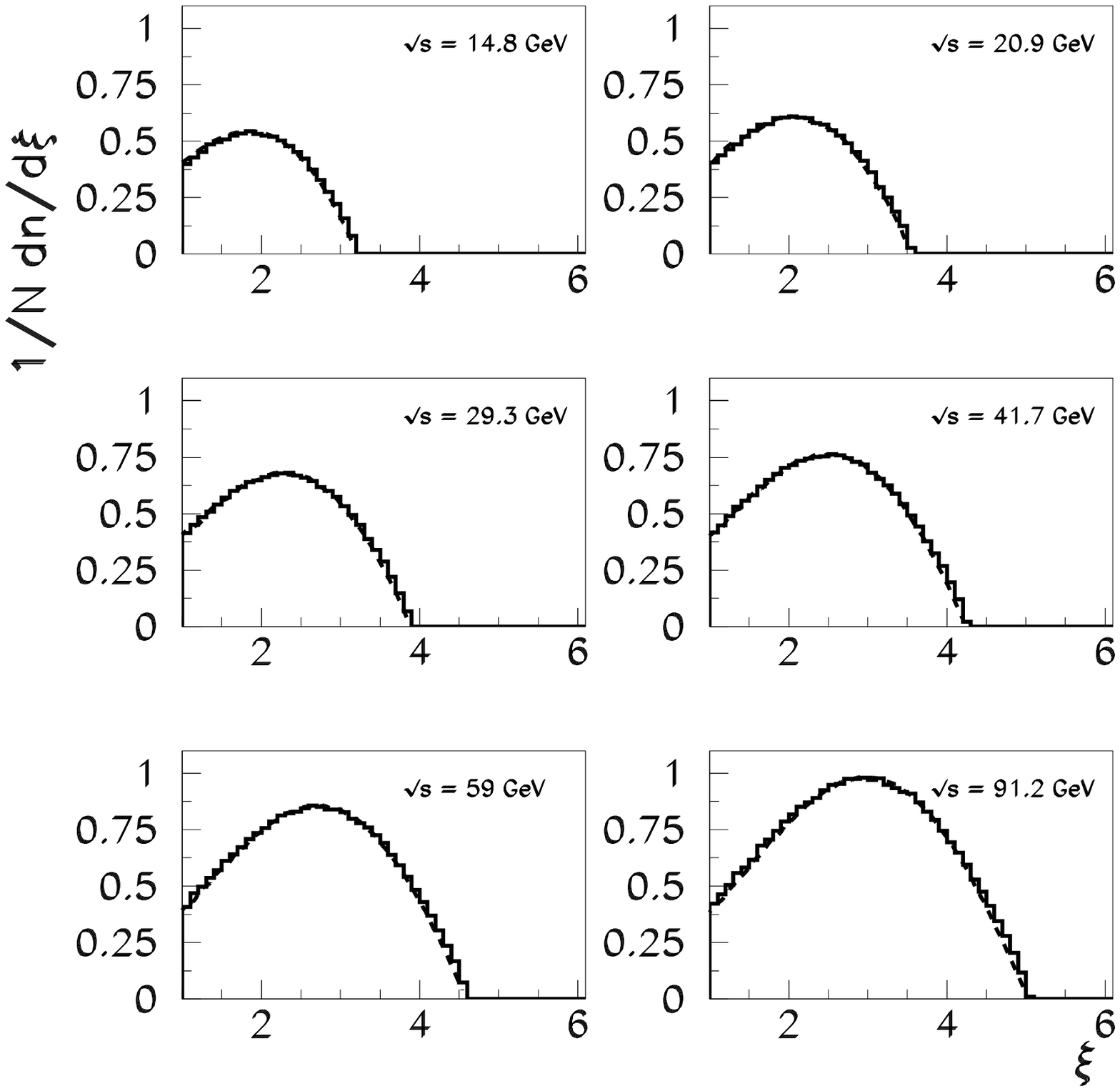,width=0.95\textwidth}}
\end{center}
\caption{ The logarithmic scaled momentum distributions, $\xi$, 
for the ARIADNE Monte Carlo generator (full line) and MLLA calculation
(dashed line) for $e^+e^-$ annihilation. 
The ARIADNE prediction has been scaled down by a factor 1.4.}
\label{fig:ee}
\end{figure}

Before investigating the evolution of the shower in DIS, the evolution
in the simpler case of $q\bar q$ pair production in $e^+e^-$ annihilation
was studied.
The spectra for both MLLA and ARIADNE were generated with a
$\Lambda=150$ MeV and a cut-off $Q_0 = 2\Lambda = {\rm PARA(3) =
PARA(28)}.$  Below $\xi \approx 1 $ there are instabilities in the
numerical integration of equation~\ref{eqn:MLLA} so all subsequent
comparisons are for $\xi > 1.$
Except for an overall normalisation discrepancy (a factor of 1.4
greater parton multiplicity in ARIADNE) 
the $\xi$ spectra are in very good agreement as illustrated
in Figure~\ref{fig:ee}.
This normalisation discrepancy is constant, independent of the $\sqrt{s}$ 
at which the events were generated at.
There is a slight tendency for the MLLA calculation to fall off quicker at
large values of $\xi$ than the ARIADNE predictions. 

\begin{table}[hbt!]
\begin{center}
\begin{tabular}{|c|c|}
\hline
 $Q$ & $x$ \\
\hline
$14.8$ & $5\times 10^{-3}$ \\
$20.9$ & $ 1\times 10^{-2}$ \\
$29.3$ & $5 \times 10^{-2}$ \\
$41.7$ & $5 \times 10^{-2}$ \\
$59.1$ & $ 0.1 $ \\
$91.2$ & $ 0.2 $ \\
\hline
\end{tabular}
\caption{ The ($x$,$Q$) analysis bins.}
\label{tab:bins}
\end{center}
\end{table}

Using LEPTO~\cite{LEPTO} to generate the electroweak cross section and 
colour flow configuration for DIS, ARIADNE was then
used to generate the subsequent QCD cascade. The event was boosted to
the Breit frame and those partons in the current fragmentation region
selected. The DIS events were generated with  fixed kinematics
that are accessible in the HERA regime. The
corresponding value of Bjorken-$x$ with $Q$ are shown in
Table~\ref{tab:bins}.
Using the same values of $\Lambda$ and $Q_0$ that was used for the
$e^+e^-$ annihilation study, the MLLA prediction is again compared to
the ARIADNE generated spectra.

\begin{figure}[pt!]
\begin{center}\mbox{\epsfig{file=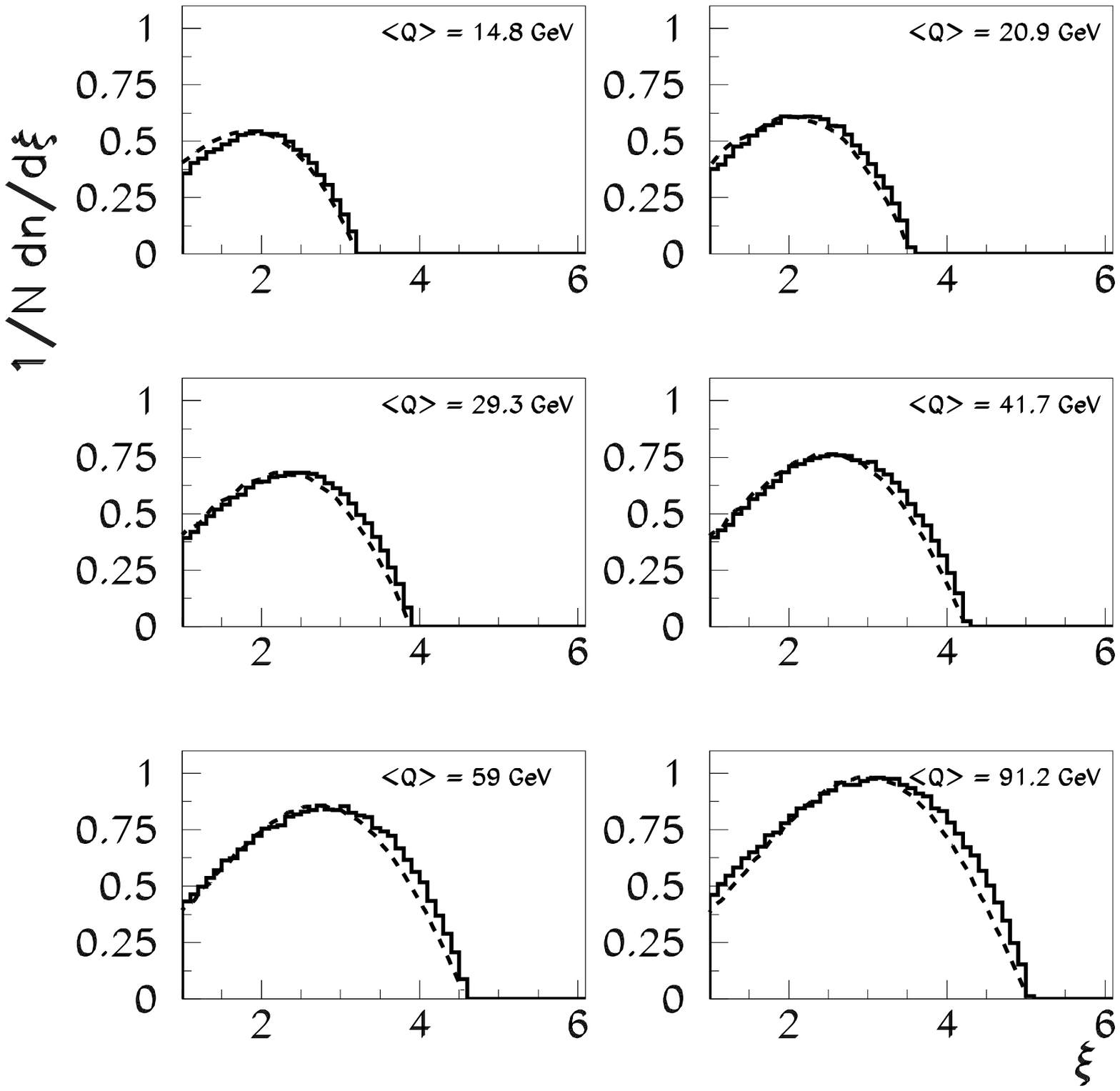,width=0.95\textwidth}}
\end{center}
\caption{ The logarithmic scaled momentum distributions, $\xi$, 
for the default ARIADNE Monte Carlo generator (full line) and MLLA calculation
(dashed line) for DIS. The distributions have been normalised to the
peak height of the MLLA calculation.}
\label{fig:dis}
\end{figure}

Figure~\ref{fig:dis} shows the default version of ARIADNE for DIS compared
to the MLLA predictions. As $Q^2$ increases the discrepancy between 
ARIADNE and the MLLA calculations becomes more pronounced. The
$\xi$ distribution  of ARIADNE peaks at higher values than the MLLA
calculation. In addition, the MLLA calculations are narrower than the
ARIADNE predictions. Again the parton multiplicity of the two
distributions are different. Unlike the $e^+e^-$ situatuion this
normalisation factor seems to exhibit a $Q$ dependence. At low $Q$
the height of the peak for ARIADNE compared to MLLA
 is a factor of 1.1 higher whilst in the highest $Q$ bin it is 1.1 lower.

\begin{figure}[pt!]
\begin{center}\mbox{\epsfig{file=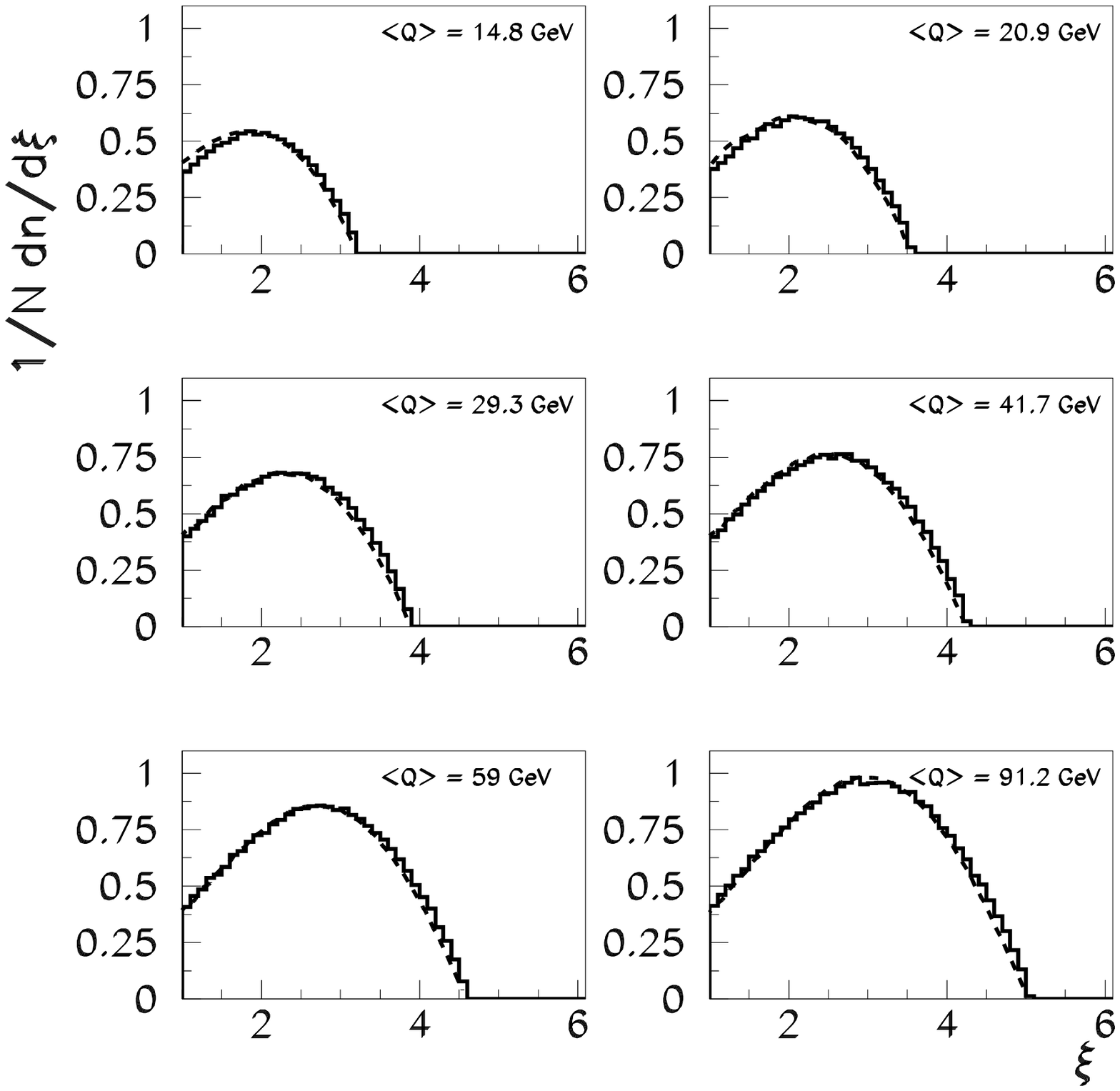,width=0.95\textwidth}}
\end{center}
\caption{ The logarithmic scaled momentum distributions, $\xi$, 
for the modified ARIADNE Monte Carlo generator (full line) and MLLA calculation
(dashed line) for DIS. The distributions have been normalised to the
peak height of the MLLA calculation.}
\label{fig:dis_mhar}
\end{figure}

In the default ARIADNE, the mechanism for soft suppression of radiation 
due to the extended source of the proton remnant results in a 
suppression of radiation in the current region of the
Breit frame at high $Q^2.$
Figure~\ref{fig:dis_mhar} shows the high $Q^2$ modified version of 
ARIADNE~\cite{leif} for DIS, where this suppression in the current
region is removed, compared to the MLLA predictions. As expected,
this modification to ARIADNE leads to a much better agreement between
the MLLA calculations and ARIADNE. The situation with the parton
multiplcity is similar. The $Q-$dependence of the ratio of the
peak heights is less than the default ARIADNE, with ARIADNE being a
factor $1.2-1.3$ higher.

In both options of the ARIADNE program there are discrepancies evident
in the lower $(x,Q)$ bins compared to the MLLA predictions. One
possible explaination of this discrepancy is given in
Ref.~\cite{eden}, where it shown that high $p_T$ emissions in DIS can
lead to the situation where the current region of the Breit frame is
depopulated. 

\section{Conclusions}
The QCD cascade as implemented in the ARIADNE Monte Carlo program is in
good agreement with the shape of the
MLLA prediction in the simple scenario of $q\bar q$ production 
in $e^+e^-$ annihilation. In the more complex situation of DIS
the agreement is not as good, unless account is taken of the additional
suppression introduced into the model in the current fragmentation region
caused by the suppression of phase space due to the extended nature of the
proton remnant.

\newpage

\newpage

\newpage

\end{document}